\documentclass[aps,prr,twocolumn,floatfix,superscriptaddress]{revtex4}

\usepackage{amssymb,amsmath,amstext}                %%   American Physical Society math etc extensions
\usepackage{graphicx}                                               %%   Include figure files                                            %%   Manage links
\usepackage{epstopdf}                                               %%   help with eps -> pdf 
\usepackage{color}                                                     %%   change color
\usepackage{bm}                                                        %%   bold math                                    
\usepackage{appendix}                                              %%   formating appendicies
\usepackage[utf8]{inputenc}
\usepackage{bbold}
\usepackage{bbm}   
\usepackage{ulem}
\usepackage{latexsym}
\usepackage[colorlinks=true,citecolor=blue,linkcolor=magenta]{hyperref}
\usepackage{mathtools}
\usepackage{mathtools}

\def\be{\begin{equation}}
\def\ee{\end{equation}}
\def\bea{\begin{eqnarray}}
\def\eea{\end{eqnarray}}
\def\ra{\rangle}
\def\la{\langle}
\def\bi{\begin{itemize}}
\def\ei{\end{itemize}}
\def\ben{\begin{enumerate}}
\def\een{\end{enumerate}}

\definecolor{dgreen} {RGB}{0,100,0}

\begin{document} 

\title{Response to comment on "Lack of a genuine time crystal in a chiral soliton model" by \"Ohberg and Wright}

\author{Andrzej Syrwid} 
\affiliation{Department of Theoretical Physics, The Royal Institute of Technology, Stockholm SE-10691, Sweden}
\affiliation{Instytut Fizyki Teoretycznej, 
Uniwersytet Jagiello\'nski, ulica Profesora Stanis\l{}awa \L{}ojasiewicza 11, PL-30-348 Krak\'ow, Poland}
\author{Arkadiusz Kosior} 
\affiliation{Instytut Fizyki Teoretycznej, 
Uniwersytet Jagiello\'nski, ulica Profesora Stanis\l{}awa \L{}ojasiewicza 11, PL-30-348 Krak\'ow, Poland}
\affiliation{Max-Planck-Institut f\"ur Physik Komplexer Systeme,
N\"othnitzer Strasse 38, D-01187, Dresden, Germany}
\author{Krzysztof Sacha} 
\affiliation{Instytut Fizyki Teoretycznej, 
Uniwersytet Jagiello\'nski, ulica Profesora Stanis\l{}awa \L{}ojasiewicza 11, PL-30-348 Krak\'ow, Poland}

\begin{abstract}
In the paper [Phys. Rev. Research {\bf 2}, 032038] we have analyzed a chiral soliton model and shown that despite the claim of \"Ohberg and Wright [Phys. Rev. Lett. {\bf 124}, 178902], there is no indication that a genuine quantum time crystal can be observed in the system. Here, we response to the recent comment on our paper written by \"Ohberg and Wright.
\end{abstract}

\date{\today}

\maketitle

%%%%%%%%%%%%%%%%%%%%%%%%%%%%%%%%%%%%%%%%%%%%%%%%%%%%%%%%%%%%%%%%%%%%%%%%%%%%%%%%%%%%%%%

A genuine quantum time crystal would be a system which would reveal periodic evolution in its lowest energy state. Quantum time crystals were proposed by Wilczek who considered attractively interacting bosons on a one-dimensional ring \cite{Wilczek2012}. In the presence of a magnetic-like flux $\alpha$ penetrating the ring, bosons were expected to form a bright soliton wavepacket that would move  periodically along the ring even if the energy of the system was minimal. It turned out that it was not possible because in the limit of the number of bosons $N\rightarrow\infty$, the lowest energy solution was  always  stationary \cite{Bruno2013b,Syrwid2017,SachaTC2020}. \"Ohberg and Wright (OW) considered the mean-field description of a similar Bose system - but in the presence of a density-dependent gauge potential - which hosts chiral solitonic solutions \cite{Ohberg2019}. They claimed that, contrary to the original Wilczek's model,  a chiral soliton model could reveal a genuine time crystal behavior. There is already a debate in the literature which can be shortly described as follows:
\ben
\item
The conclusion based on the mean-field results presented in the initial article by OW \cite{Ohberg2019} was not correct
due to an erroneous expression for the energy of the system in the laboratory frame employed by OW,  which we pointed out in Ref.~\cite{SyrwidKosiorSacha2020} and which was admitted in Ref.~\cite{ReplyOhbergWright2020}.
\item
However, in their response \cite{ReplyOhbergWright2020}, OW still claimed that a genuine time crystal could be observed in the system due to the quantization of a chiral soliton's velocity  if a soliton was not represented by a strongly localized wavepacket on a ring. In Ref.~\cite{sksPRR2020}, we have shown that for the mean-field equations OW consider, there exists a larger class of solutions where solitons can move with any velocity and the laboratory frame energy is always minimized by a stationary soliton solution.  Consequently there is no evidence for a genuine time crystal.
\item
In the present comment \cite{OWcomment2020}, OW sustain that a genuine time crystal can be observed in the system if the limit $N\rightarrow\infty$ is not taken. Their claim is supported by no evidence. 
\een

\begin{figure}
\includegraphics[width=1\columnwidth]{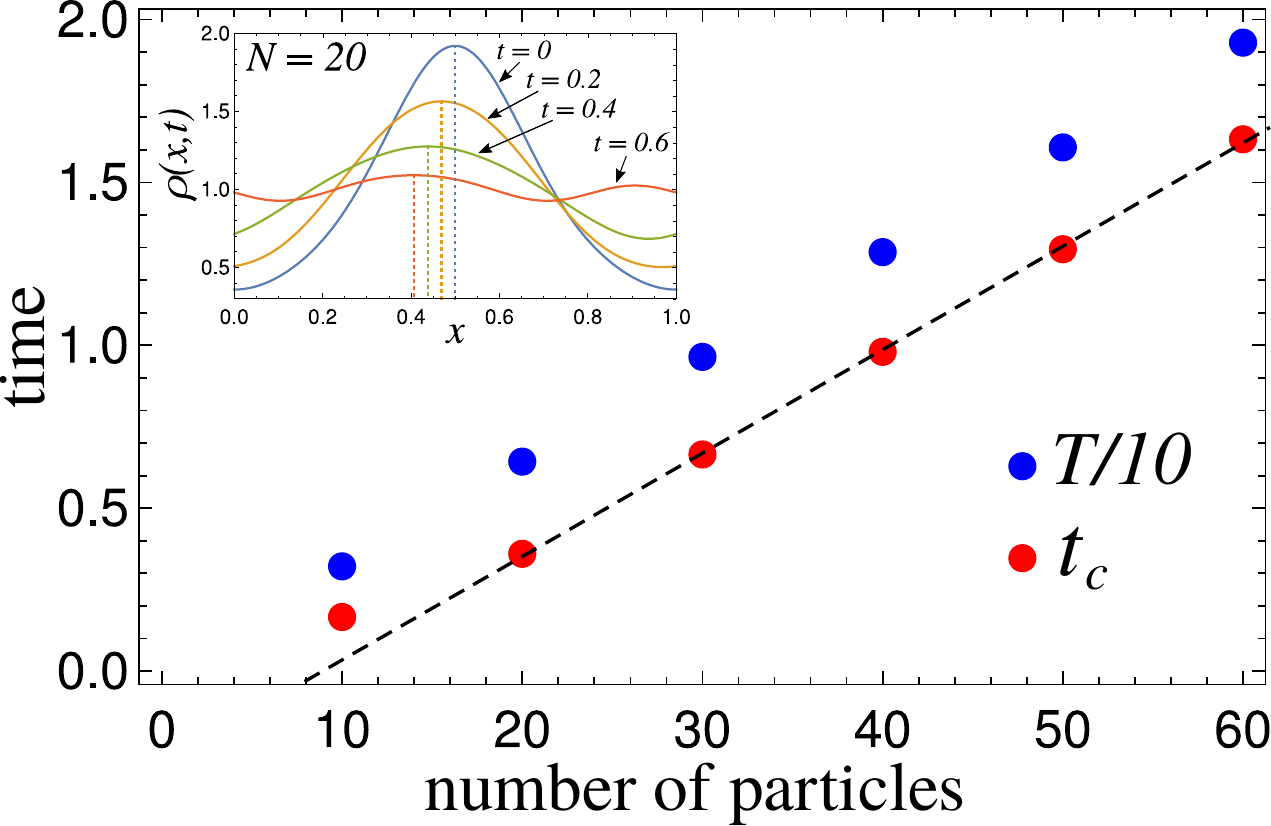} 
\caption{
Comparison between the lifetime $t_c$ and the period $T=N/0.99\pi$ of the soliton's motion around the ring for different numbers of particles $N$. Note that $t_c$ (red dots) increases linearly with $N$ (dashed black line represents the linear fit) and is more than 10 times shorter than the period $T$ (note that blue dots correspond to $T/10$). For illustration, in the inset we present how the soliton structure visible in $\rho(x,t)$ dies out in time for $N=20$. Vertical dotted lines guide the eye and indicate the position of the soliton clump in four different time moments $t=\{0,0.2,0.4,0.6\}$. 
  }
\label{fig_1}   
\end{figure}

In all mentioned papers, Refs.~\cite{Ohberg2019,ReplyOhbergWright2020,OWcomment2020}, OW describe the Bose system within the mean-field theory which implicitly assumes the $N\rightarrow\infty$ limit. The mean-field equations possess chiral solitonic solutions. Regardless of whether the solitons are or are not strongly localized on a ring, there is no quantization rule for the soliton velocity. All soliton solutions found minimize the energy if they do not move. The large $N$ limit is necessary in order to ensure that a soliton lives forever. For $N<\infty$, it is in principle possible to have a periodic evolution of a soliton when we start from the ground state,  but its lifetime cannot be infinite due to the quantum fluctuations \cite{Syrwid2017}. In Ref.~\cite{OWcomment2020} OW claim that the lifetime of a soliton for $N<\infty$ can be long enough to observe the time crystalline behavior, but they provide no evidence that it is true. Their arguments concerning the non-zero flux of bosons along the ring in the ground state of the system for $N<\infty$ can be also applied to the original Wilczek's model \cite{Wilczek2012}, where the lowest energy state corresponds to the total momentum $P=2\pi n$ with $n\in\mathbb{Z}$ minimizing the center of mass kinetic energy $(P-\alpha N)^2/(2N)$ (assuming the unit ring circumference and $m=\hbar=1$). When the interparticle attraction is strong enough, then a bright soliton forms spontaneously and is expected to propagate with a velocity $P/N-\alpha$, which in the ground state can be nonzero for $N<\infty$ only \cite{Syrwid2017}. The crucial question is whether for $N<\infty$ the period of the soliton motion  along the ring is shorter than the time needed to spread the soliton along the ring due to the quantum many-body effects. To address this question it is sufficient to consider $\alpha N  \in(-\pi,\pi)$, for which the ground state corresponds to $n=0$. Note that the soliton velocity is maximized when the magnetic-like flux $|\alpha| \rightarrow \pi/N$. Thus, following the idea presented in Ref.~\cite{Syrwid2017}, we analyze a density-density correlation function $\rho_2(x_2,t_2;x_1,t_1)\propto \la \hat\psi^\dagger(x_1,t_1) \hat\psi^\dagger(x_2,t_2)\hat\psi(x_2,t_2)\hat\psi(x_1,t_1)\ra$ for $\alpha N = 0.99\pi$ and the strength of the attractive interactions between particles given by $g_0(N-1)=-12.5$, where $\hat\psi$ is the bosonic field operator and $g_0$ denotes the strength of the attractive two-body contact interactions. Note that in the mean-field description a bright soliton on a ring appears in the ground state for $g_0(N-1)<-\pi^2$ (see for example Refs.~\cite{Carr2000,Syrwid2020}). Therefore our choice of parameters 
 guarantees a formation of a bright soliton wavepacket that is not strongly localized  and propagates with 99\% of the maximal velocity if we start with the ground state of the system consisting of $N<\infty$ bosons. To break the space translation symmetry possessed by the ground state of the translationally invariant system we perform an initial measurement of a single boson at $x_1=0.5$ and at $t_1=0$. Thanks to this, we can monitor a temporal behaviour of the single particle density after the initial measurement, $\rho(x,t)=\rho_2(x,t;0.5,0)$, which reveals a soliton-like wavepacket propagating along the ring with the expected velocity $-0.99\pi/N$ and decaying in time due to quantum many-body effects. The lifetime $t_c$ of the soliton can be estimated basing on the contrast quantity $\mathcal{C}(t)=\{\mathrm{max}_x[\rho(x,t)] - \mathrm{min}_x[\rho(x,t)]\}/\{\mathrm{max}_x[\rho(x,t)] + \mathrm{min}_x[\rho(x,t)]\}$ (see also Ref.~\cite{Syrwid2017}). Here we define $t_c$  as the minimal time for which $\mathcal{C}(t_c)\approx 0.5\mathcal{C}(0)\approx 0.34$. In Fig.~\ref{fig_1} we show that in the Wilczek's model regardless of how small $N$ one chooses, it is not possible to observe even a single revolution of a soliton before it spreads along the ring. As there is no evidence that the situation is different in the case of the chiral soliton model, there is no argument to claim that a time crystal dynamics can be observed in the system.

%%%%%%%%%%%%%%            
%%%%%%%%%%%%%% 

\section*{Acknowledgements}
Support of the National Science Centre, Poland via Projects No.~2018/28/T/ST2/00372 (A.S.) and No.~2018/31/B/ST2/00349
(A.K. and K.S.) is acknowledged. 
%%%%%%%%%%%%%%%%%%%%%%%%%%%%%%%%%%%%%%%%%%%%%%%Supported%%%%%%%%%%%%%%%%%%%%%%%%%%%%%%%%%%%%%%%%

\bibliography{ref_tc_book}

%%%%%%%%%%%%%%%%%%%%%%%%%%%%%%%%%%%%%%%%%%%%%%%%%%%%%%%%%%%%%%%%%%%%%%%%%%%%%%%%%%%%%%%

\end{document}